\documentclass[12pt]{article}
\usepackage{amssymb}
\usepackage{color,graphicx}
\usepackage{amsmath}

\newcommand{\be}{\begin{equation}}
\newcommand{\ee}{\end{equation}}

\newcommand{\ba}{\begin{eqnarray}}
\newcommand{\ea}{\end{eqnarray}}

\title{Quantum cosmology of the multi-field scalar matter: some exact solutions}
\author{A.A.Andrianov\footnote{V.A. Fock Department of Theoretical Physics, Saint-Petersburg State University, ul. Ulianovskaya, 198504 St. Petersburg, Russia}, O.O. Novikov\footnotemark[1], Chen Lan\footnotemark[1]}

\date{}

\begin{document}

\maketitle

\begin{abstract}
We consider the gravity interacting with matter scalar fields and quantized in the minisuperspace approach in which the wave functional is described by the Wheeler-DeWitt equations (WdW). Assuming the domination of the homogeneous and isotropic geometry the leading contributions to the wave functional in the approximation of the minisuperspace with Friedmann-Robertson-Walker metric (FRW) and spatially uniform scalar fields are considered. The model of several scalar fields with exponential potentials and kinetic terms admitting such a special mixing that ultimately it is possible to separate the variables in the WdW equation and to find its exact solution in terms of the special functions is proposed. The semiclassical approximation is thoroughly investigated and the boundary conditions permitting the physical solution selection for classical cosmologies are chosen.
\end{abstract}

\section{Introduction}
The exact Einstein equation solutions in the presence of the matter play important role in the cosmology. Last decades the relativistic models with the scalar matter that are applicable to describe the inflationary expansion in the early stages of the evolution \cite{inflation1},\cite{inflation2} and to describe the cosmic acceleration \cite{cosmic1},\cite{cosmic2} at later stages (the quintessence and the dark matter models \cite{dark1}--\cite{dark5}) have been quite popular. At the same time there are only few known exactly solvable cosmological models with scalar matter. One such a model is of a flat Friedmann universe filled with a scalar field with the exponential potential. Various solutions for one scalar field were studied in \cite{power-law}--\cite{power-law5}. Recently the general Einstein equation solution for this model were found \cite{general0}--\cite{general9}.

We mentioned classical solutions for the universe geometry evolution. However to understand in the details the metric singularity in the moment of the Big Band one has to know the quantum properties of the gravity interacting with the quantum matter fields. Without a doubt any exact solution for the quantum univers is invaluable even in the restricted sense - with few degrees of freedom quantized. To ensure unambigous classical limit the ADM quantization in the minisuperspace is commonly used practical essence of which is given by the Wheeler-DeWitt equations (see \cite{kiefer} and references therein) reflecting the constraints due to the symmetry under diffeomorphisms. Generally speaking these functional equations are very hard to solve. However to some extent one can assume that for approximately flat universes the leading quantum effects are determined by the spacially uniform and isotropic fluctuations of the gravity and matter whereas the inhomogeneities and anisotropies can be succesfully analyzed in the approximation of the weak quantum fluctuations. In this case we come to the minisuperspace reduction with gravity and matter fields depending only on time. As a matter of fact such a model is a quantum mechanics with finite number of the degrees of freedom.

In this work we assume such a domination of the spatially uniform isotropic geometry and study the leading contribution to the wave functional in the FRW metric minisuperspace approximation with auxiliary lapse field that provides the time reparametrization invariance and scalar fields depending only on time. The model of several scalar fields with kinetic terms admitting mixing is proposed. This mixing is chosen in such a way that ultimately it is possible to separate the variables in the WdW equation and to find the exact solutions in the terms of special functions. In the section 2 we study the classical Einstein-Friedmann euqations, solve them and thus provide the basis for consistent quantization of the WdW equation in the minisuperspace approximation (section 3). Exact quantum solution of this equation are obtained in the section 4. In the same section the boundary conditions for selection of the physical quantum solutions that admit reasonable classical cosmologies are introduced. Semiclassical (WKB) approximation is thoroughly investigated in the section 5. Future plans and prospects of this class of models are briefly discussed in the conclusion.

\section{Model of several scalar fields with exponential potentials}

Let us consider the following model with several scalar fields minimally coupled to the gravity,
\begin{equation}
S=\int d^4x \sqrt{-g}\Big(-\frac{1}{2\kappa^2}R+\sum_{ab} \frac{1}{2}M_{ab}\partial_\mu\phi_a\partial^\mu\phi_b-\sum_a V_a e^{\lambda_a\phi_a}\Big)
\end{equation}
where index $a$ goes from $1$ to $n$.

Let us restrict our consideration to the minisuperspace corresponding to Friedmann-Robertson-Walker metrics and spatially uniform fields,
\begin{equation}
ds^2=N^2(t)dt^2-e^{2\rho(t)}d\vec{x}^2,\quad\phi_a=\phi_a(t).
\end{equation}
The corresponding minisuperspace action takes the form,
\begin{equation}
S=\int dt\,e^{3\rho}\Big(-\frac{3}{\kappa^2}\frac{\dot{\rho}^2}{N}+\sum_{ab}M_{ab}\frac{\dot{\phi}_a\dot{\phi}_b}{2N}-N\sum_aV_a e^{\lambda_a\phi_a}\Big).
\end{equation}

It is convenient to search for the classical solutions in the Hamilton formulation. The canonical momenta are,
\begin{equation}
p_\rho=-\frac{6}{\kappa^2}e^{3\rho}\frac{\dot{\rho}}{N},\quad p_a=e^{3\rho}\sum_b M_{ab}\frac{\dot{\phi}_b}{N},\quad p_N=0.
\end{equation}
As always in the theory of gravity, the Hamiltonian becomes purely a constraint with a Lagrangian multiplier $N$,
\begin{equation}
H=Ne^{-3\rho}\left[-\frac{\kappa^2}{12}p_\rho^2+\frac{1}{2}\sum_{ab}(M^{-1})_{ab}p_ap_b+\sum_a V_a e^{6\rho+\lambda_a \phi_a}\right].\label{hamilton1}
\end{equation}
To eliminate the dependence of the exponent term on the different degrees one should to perform the following canonical transformation,
\begin{equation}
\begin{pmatrix}\rho, p_\rho\\ \phi_a, p_a\end{pmatrix}\mapsto\begin{pmatrix}\rho, \omega\\ \chi_a, \pi_a\end{pmatrix},\label{cantrans1}
\end{equation}
\begin{equation}
\omega=p_\rho-\sum\limits_a\frac{6}{\lambda_a}p_a,\quad \chi_a=\lambda_a\phi_a+6\rho,\quad \pi_a=\frac{1}{\lambda_a}p_a,\label{cantrans2}
\end{equation}
then the Hamiltonian takes the form,
\begin{eqnarray}
&H=Ne^{-3\rho}\Bigg[&-\frac{\kappa^2}{12}\omega^2-\kappa^2\omega\sum_a\pi_a+\sum_{ab}\Bigl(\frac{\lambda_a\lambda_b}{2}(M^{-1})_{ab}-3\kappa^2\Bigr)\pi_a\pi_b\nonumber\\
&&+\sum_a V_a e^{\chi_a}\Bigg].
\end{eqnarray}

Note that new canonical momentum $\omega$ is conserved on the constraint surface, i.e.
\begin{equation}
\{\omega,H\}=-3H\approx 0
\end{equation}

Generally speaking this model is not exactly solvable in the case of several fields. Let us assume the special form of the kinetic term matrix,
\begin{equation}
(M^{-1})_{ab}=\frac{D_a\delta_{ab}+6\kappa^2}{\lambda_a\lambda_b},\label{matrixshift}
\end{equation}
such that the kinetic energy of the $\chi_a$ fields becomes diagonalized,
\begin{equation}
H=Ne^{-3\rho}\left[-\frac{\kappa^2}{12}\omega^2-\kappa^2\omega\sum_a\pi_a+\frac{1}{2}\sum_a D_a\pi_a^2+\sum_a V_a e^{\chi_a}\right].
\end{equation}
As will be shown below this permits separation of variables in the classical as well as in the quantum case.

The classical solutions can easily be found by solving Hamiltonian equations of motion. Let's choose the gauge,
\begin{equation}
N=e^{3\rho}.
\end{equation}
The equations of motion on the constraint surface take the form,
\begin{eqnarray}
\dot{\chi}_a=\{\chi_a,H\}\simeq-\kappa^2\omega+D_a\pi_a,&&\dot{\pi}_a=\{\pi_a,H\}\simeq-V_ae^{\chi_a},\\
\dot{\rho}=\{\rho,H\}\simeq -\frac{\kappa^2}{6}\omega-\kappa^2\sum_a\pi_a,&&\dot{\omega}=\{\chi_a,H\}\simeq0,
\end{eqnarray}

Taking the derivative with respect to time of the first equation we get,
\begin{equation}
\ddot{\chi}_a=-D_aV_ae^{\chi_a},
\end{equation}
from which the following solution is obtained,
\begin{equation}
e^{\chi_a}=\frac{2P_a^2}{D_aV_a}\frac{1}{\cosh^2(P_at-Q_a)},\label{classsol1}
\end{equation}
where $P_a$,$Q_a$ - integration constants. Substituting this result into other equations we obtain general solution,
\begin{eqnarray}
&\pi_a=-\frac{2P_a}{D_a}\tanh(P_at-Q_a)+\frac{\kappa^2}{D_a}\omega,\quad\rho=\rho_0+\sum\limits_a \rho_a,\\
&\rho_a=-\frac{4P_a^2}{D_a}\omega t+\frac{\kappa^2}{D_a}\ln\cosh^2(P_at-Q_a),\\
&P_a^2=\frac{\kappa^2\omega^2}{24}(6\kappa^2+D_an^{-1}+C_aD_a),\quad \sum\limits_a C_a=0.\label{classsol2}
\end{eqnarray}
where some integration constants happen to be fixed thanks to the constraint equation $H=0$.

Note that for solutions to be real requires,
\begin{equation}
\frac{P_a^2}{D_aV_a}>0,
\end{equation}
which constrains the classically allowed values of $C_a$. For $D_aV_a<0$ this condition demands $P_a^2=-\tilde{P_a}^2<0$,
\begin{eqnarray}
&&e^{\chi_a}=\frac{2\tilde{P}_a^2}{|D_aV_a|}\frac{1}{\cos^2(\tilde{P}_at-\tilde{Q}_a)},\\
&&\rho_a=+\frac{4\tilde{P}_a^2}{D_a}\omega t+\frac{\kappa^2}{D_a}\ln\cos^2(\tilde{P}_at-\tilde{Q}_a).
\end{eqnarray}
For $P_at-Q_a=\pm\frac{\pi}{2}$ the cosine vanishes and both $\chi_a$ and $\rho$ tend to infinity. Hence in this case the solution is to be considered on the finite interval of $t$. Note that in the case of $P_a^2>0$,$D_aV_a>0$ the classically allowed region of $\chi_a$ is bounded from above while for $P_a^2$, $D_aV_a<0$ it is bounded from below.

The $D_aV_a<0$ case may correspond to the phantom fields (quintoms \cite{quintom}) or negative potentials unbounded from below. The question of the stability for such cosmological systems lays beyond the scope of this paper and will be left for the further analysis. Nevertheless it should be noted that according to \eqref{matrixshift} sufficiently small negative value $D_a$ may still correspond to the positive-definite kinetic term matrix $M_{ab}$.

Let us consider the case of one field with the different time variable analogous to the one considered in \cite{general8},\cite{general8_1}.
\begin{equation}
\frac{d\tau}{dt}=\frac{P}{DV}e^{\frac{\chi}{2}},
\end{equation}
For $P^2>0$,$DV>0$ we get,
\begin{eqnarray}
\tau=\tau_0+\arctan\tanh\frac{Pt-Q}{2},\quad
e^{\chi_a}=\frac{2P_a^2}{D_aV_a}\frac{1}{\cos 2(\tau-\tau_0)},\\
\rho_a=-\frac{4P_a}{D_a}\omega (Q+2\operatorname{arctanh}\tan(\tau-\tau_0)) +\frac{\kappa^2}{D_a}\ln\cos 2(\tau-\tau_0),
\end{eqnarray}
Note that original time variable $t$ corresponds only to the finite interval of $\tau$. However at the ends of these interval the metric becomes singular $e^{2\rho}=0$ and beyond it becomes complex.

For $P^2<0$,$DV<0$,
\begin{eqnarray}
\tau=\tau_0+\operatorname{arctanh}\tan\frac{\tilde{P}t-\tilde{Q}}{2},\quad
e^{\chi_a}=\frac{2\tilde{P}_a^2}{|D_aV_a|}\frac{1}{\cosh 2(\tau-\tau_0)},\\
\rho_a=\frac{4\tilde{P}_a}{D_a}\omega (\tilde{Q}+2\arctan\tanh(\tau-\tau_0)) +\frac{\kappa^2}{D_a}\ln\cosh 2(\tau-\tau_0),
\end{eqnarray}
I.e. the finite interval of $t$ in this case corresponds to the whole real axis of $\tau$.

Cosmic time for which the metrics takes the form,
\begin{equation}
ds^2=dt_{cosm}^2-e^{2\rho}d\vec{x}^2,\quad t_{cosm}=\int_{t_0}^t dt\, N.
\end{equation}

Let us consider the case with all $D_a>0$,$V_a>0$ assuming that all fields pass the turning point approximately at the same time i.e. $Q_a\simeq 0$. For large values of $|t|$ the metric factor $\rho$ behaves as a linear function,
\begin{equation}
\rho\sim \rho_0-\Big[\frac{\kappa^2}{6}+\sum_a\frac{\kappa^4}{D_a}\Big]\omega^3 t+\Big[\sum_a\frac{\kappa^3}{2\sqrt{6}D_a}\sqrt{6\kappa^2+D_an^{-1}+C_aD_a}\Big]\omega|t|
\end{equation} 
Let $\omega<0$. Then the cosmic time between the Big Band at $t\rightarrow-\infty$ and the turning point at $t=0$ is finite. On the other hand the cosmic time between the turning point and the Big Crunch at $t\rightarrow+\infty$ is finite only if,
\begin{equation}
\omega^2<\frac{\sum_a\frac{\kappa\sqrt{6}}{2D_a}\sqrt{6\kappa^2+D_an^{-1}+C_aD_a}}{1+\sum_a\frac{6\kappa^2}{D_a}}
\end{equation}
To consider the solution with $\omega>0$ it is sufficient to reflect the time $t\mapsto -t$ in the solution with negative $\omega$.

If for at least one field $D_a<0$, $V_a>0$ then the finite interval of $t$ for which the solution is defined corresponds to the infinite cosmic time between the Big Bang and the Big Crunch.

\section{Wheeler-DeWitt equation}

The Wheeler-DeWitt equation for the minisuperspace under consideration is obtained by canonical quantization for the Hamilton function \eqref{hamilton1},
\begin{equation}
Ne^{-3\rho}\left[\hbar^2\frac{\kappa^2}{12}\partial_\rho^2-\hbar^2\frac{1}{2}\sum_{ab}(M^{-1})_{ab}\partial_a\partial_b+\sum_a V_a e^{6\rho+\lambda_a \phi_a}\right]\Psi(\rho,\{\phi_a\})=0.
\end{equation}
Factor $Ne^{-3\rho}$ will not play the role in finding the solutions and thus it will be ommited.

Canonical transformation \eqref{cantrans1},\eqref{cantrans2} in the quantum case becomes simply the coordinate transformation and the corresponding transformation of the partial derivatives after which the equation takes the form,
\begin{eqnarray}
&&\Biggl[\hbar^2\frac{\kappa^2}{12}\partial_\rho^2+\hbar^2\kappa^2\partial_\rho\sum_a\partial_a-\hbar^2\sum_{ab}\Bigl(\frac{\lambda_a\lambda_b}{2}(M^{-1})_{ab}-3\kappa^2\Bigr)\partial_a\partial_b\nonumber\\
&&\quad+\sum_a V_a e^{\chi_a}\Biggr]\Psi(\rho,\{\chi_a\})=0.\label{WdW}
\end{eqnarray}

Because now there is no $\rho$ in the potential, it is convenient to perform the Fourier transform,
\begin{equation}
\Psi(\rho,\{\chi_a\})=\int d\omega\, e^{\frac{i}{\hbar}\omega\rho} \Psi(\omega,\{\chi_a\}),
\end{equation}
\begin{eqnarray}
&&\Biggl[-\frac{\kappa^2}{12}\omega^2+i\hbar\kappa^2\omega\sum_a\partial_a-\hbar^2\sum_{ab}\Bigl(\frac{\lambda_a\lambda_b}{2}(M^{-1})_{ab}-3\kappa^2\Bigr)\partial_a\partial_b\nonumber\\
&&\quad+\sum_a V_a e^{\chi_a}\Biggr]\Psi(\omega,\{\chi_a\})=0
\end{eqnarray}

In the case of the special form of the kinetic term matrix
\begin{equation}
(M^{-1})_{ab}=\frac{D_a\delta_{ab}+6\kappa^2}{\lambda_a\lambda_b},
\end{equation}
mixed terms are canceled and partial solutions can be obtained by separation of variables,
\begin{equation}
\Psi(\omega,\{\chi_a\})=\prod_a \Psi_a(\omega,\chi_a),
\end{equation}
with $\Psi_a$ depending only on one coordinate $\chi_a$.

The corresponding equations take the form,
\begin{equation}
\Big[-\frac{\kappa^2}{12}\omega^2(n^{-1}+C_a)+i\hbar\kappa^2\omega\partial_a-\hbar^2\frac{D_a}{2}\partial_a^2+V_ae^{\chi_a}\Big]\Psi_a(\omega,\chi_a)=0,\label{eqsepvar}
\end{equation}
where $C_a$ are arbitrary constants under condition $\sum\limits_a C_a=0$.

\section{Exact solutions of WdW equation}

Let us consider now the exact solutions of the equations \eqref{eqsepvar}. After the variable change,
\begin{equation}
\Psi=e^{\frac{i}{\hbar}\frac{\kappa^2\omega}{D_a}\chi_a} f,\quad z=\frac{2}{\hbar}\sqrt{\frac{2V_a}{D_a}}e^{\chi_a/2},
\end{equation}
we obtain the modified Bessel equation,
\begin{equation}
\Big[z^2\partial_z+z\partial_z+\nu_a^2-z^2\Big]f(\omega,z)=0,
\end{equation}
\begin{equation}
\nu_a=\frac{4P_a}{\hbar D_a},\quad P_a^2=\frac{\kappa^2\omega^2}{24}(6\kappa^2+D_an^{-1}+C_aD_a),
\end{equation}
with $P_a$ expressed similarly to the classical solution \eqref{classsol1},\eqref{classsol2},

Therefore we obtain the exact solution,
\begin{eqnarray}
\Psi_a(\omega,\chi_a)=A_\omega^{(+)}\Psi_a^{(+)}(\omega,\chi_a)+A_\omega^{(-)}\Psi_a^{(+)}(\omega,\chi_a),\\
\Psi_a^{(+)}=e^{\frac{i}{\hbar}\frac{\kappa^2\omega}{D_a}\chi_a} K_{i\nu_a}\Big(\frac{2}{\hbar}\sqrt{\frac{2V_a}{D_a}}e^{\chi_a/2}\Big),\label{WdWexactsol}\\
 \Psi_a^{(-)}= e^{\frac{i}{\hbar}\frac{\kappa^2\omega}{D_a}\chi_a}L_{i\nu_a}\Big(\frac{2}{\hbar}\sqrt{\frac{2V_a}{D_a}}e^{\chi_a/2}\Big),
\end{eqnarray}
where $L_{\nu}$ take the following form in terms of modified Bessel functions,
\begin{equation}
L_{\nu}(z)=\frac{\pi i}{2\sin(\mu\pi)}\Big(I_{\nu}(z)+I_{-\nu}(z)\Big)
\end{equation}
Let us consider the case $P_a^2>0$, $D_aV_a>0$. Consider the limit $\chi_a\rightarrow-\infty$ ($z\rightarrow 0+$) which also corresponds to $\rho\rightarrow-\infty$ for fixed $\phi_a$. Using the asymptotics of the modified Bessel functions with purely imaginary order near zero \cite{Dunster} we get that both particular solutions behave like plane waves,
\begin{eqnarray}
\Psi_a^{(+)} \sim -\sqrt{\frac{\pi}{\nu_a\sinh{\nu_a\pi}}}e^{\frac{i}{\hbar}\frac{\kappa^2\omega}{D_a}\chi_a}\left[\sin\Big(\frac{\nu_a\chi_a}{2}-\delta_a\Big)+O\Big(e^{\chi_a}\Big)\right],\\
\Psi_a^{(-)} \sim \sqrt{\frac{\pi}{\nu_a\sinh{\nu_a\pi}}}e^{\frac{i}{\hbar}\frac{\kappa^2\omega}{D_a}\chi_a}\left[\cos\Big(\frac{\nu_a\chi_a}{2}-\delta_a\Big)+O\Big(e^{\chi_a}\Big)\right],\\
\delta_a=\nu_a\ln{\frac{1}{\hbar}\sqrt{\frac{2V_a}{D_a}}}+\operatorname{arg}\Big[\Gamma(1+i\nu)\Big].
\end{eqnarray}

Let us consider now the limit $\chi_a\rightarrow +\infty$ ($z\rightarrow+\infty$) which also corresponds to $\rho\rightarrow+\infty$ for fixed $\phi_a$. Using the leading asymptotics of Bessel functions $I_nu(z)$ and $K_nu(z)$ at large values of argument $|z|\to\infty$ (according to 8.451 in \cite{Gradshtein}) we obtain,
\begin{equation}
\Psi_a^{(\pm)}\sim \frac{\pi^{\pm1/2}\sqrt{\hbar}}{2}\Big(\frac{V_a}{2D_a}\Big)^{1/4}e^{-\frac{\chi_a}{4}}\exp\Big\{\mp\frac{2}{\hbar}\sqrt{\frac{2V_a}{D_a}}e^{\frac{\chi_a}{2}}\Big\}e^{\frac{i}{\hbar}\frac{\kappa^2\omega}{D_a}\chi_a}.\label{WdWsolasym}
\end{equation}
The selection of the solution may be done based on the condition of square integrability by $\rho$ from $-\infty$ to $+\infty$ to ensure the $\pi_\rho$ operator hermiticity and the self-consistency of the Dira or BRST/BFV quantization scheme \cite{BarvKam}. Because this condition has to be satisfied for any value of $\{\phi_a\}$ it yields the decreasing of the solution at large values of any $\chi_a$. Thus for the case of $P_a^2>0$,$D_aV_a>0$,
\begin{equation}
\Psi(\rho,\{\chi_a\})=\int d\omega\, A(\omega) e^{\frac{i}{\hbar}\omega\rho} \prod_a e^{\frac{i}{\hbar}\frac{\kappa^2\omega}{D_a}\chi_a} K_{i\nu_a}\Big(\frac{2}{\hbar}\sqrt{\frac{2V_a}{D_a}}e^{\chi_a/2}\Big).
\end{equation}

Let us consider the case of $P_a^2=-\tilde{P}_a^2$,$V_aD_a<0$. Then it is convenient to write the solution in the form,
\begin{equation}
\Psi_a(\omega,\chi_a)=\tilde{A}_\omega^{(+)}\tilde{\Psi}_a^{(+)}(\omega,\chi_a)+\tilde{A}_\omega^{(-)}\tilde{\Psi}_a^{(+)}(\omega,\chi_a),
\end{equation}
\begin{equation}
\Psi_a^{(\pm)}=e^{\frac{i}{\hbar}\frac{\kappa^2\omega}{D_a}\chi_a} J_{\pm\tilde{\nu}_a}\Big(\frac{2}{\hbar}\sqrt{\Big|\frac{2V_a}{D_a}\Big|}e^{\chi_a/2}\Big),\label{WdWexactsol2}
\end{equation}
where $\tilde{\nu}_a=|\nu_a|$. Now the limit $\chi_a\rightarrow+\infty$ ($y_a=\frac{2}{\hbar}\sqrt{\Big|\frac{2V_a}{D_a}\Big|}e^{\chi_a/2}\rightarrow+\infty$) corresponds to the oscillations,
\begin{eqnarray}
\Psi_a^{(\pm)}\sim \sqrt{\frac{\hbar}{\pi}}\Big(\frac{V_a}{2D_a}\Big)^{1/4}e^{-\frac{\chi_a}{4}}\Bigg[\cos\Big(\frac{2}{\hbar}\sqrt{\Big|\frac{2V_a}{D_a}\Big|}e^{\chi_a/2}\mp\frac{\pi}{2}\tilde{\nu}_a-\frac{\pi}{4}\Big)\\
-\sin\Big(\frac{2}{\hbar}\sqrt{\Big|\frac{2V_a}{D_a}\Big|}e^{\chi_a/2}\mp\frac{\pi}{2}\tilde{\nu}_a-\frac{\pi}{4}\Big)+O(e^{-\chi_a})\Bigg],
\end{eqnarray}
whereas in the limit $\chi_a\rightarrow-\infty$ ($y\rightarrow 0+$) (using the Bessel function asymptotics from 8.440 in \cite{Gradshtein}),
\begin{equation}
\tilde{\Psi}_a^{(\pm)} \sim e^{\frac{i}{\hbar}\frac{\kappa^2\omega}{D_a}\chi_a} \Big(\frac{1}{\hbar}\sqrt{\Big|\frac{2V_a}{D_a}\Big|}\Big)^{\pm\tilde{\nu}_a} \exp\Big(\pm\frac{\nu_a\chi_a}{2}\Big).
\end{equation}
Square integrability by $\rho$ demands the selection of the decreasing solution, i.e.
\begin{equation}
\Psi(\rho,\{\chi_a\})=\int d\omega\, A(\omega) e^{\frac{i}{\hbar}\omega\rho} \prod_a e^{\frac{i}{\hbar}\frac{\kappa^2\omega}{D_a}\chi_a} J_{\tilde{\nu}_a}\Big(\frac{2}{\hbar}\sqrt{\Big|\frac{2V_a}{D_a}\Big|}e^{\chi_a/2}\Big).
\end{equation}
Note that in both cases the selection was done according to the behavious in the classically forbidden under-barrier regieon.

\section{WKB asymptotics}

Let us consider WKB asymptotics of the Wheeler-DeWitt equation \eqref{WdW},
\begin{equation}
\Psi(\rho,\{\chi_a\})=\psi(\rho,\{\chi_a\}) e^{\frac{i}{\hbar}S(\rho,\{\chi_a\})}
\end{equation}
Hamilton-Jacobi equation takes the form,
\begin{eqnarray}
&&-\frac{\kappa^2}{12}(\partial_\rho S)^2-\kappa^2(\partial_\rho S)\sum_a (\partial_a S)\nonumber\\
&&+\sum_{ab}\Bigl(\frac{\lambda_a\lambda_b}{2}(M^{-1})_{ab}-3\kappa^2\Bigr)(\partial_a S)(\partial_b S)+\sum_a V_a e^{\chi_a}=0.\label{HJeq}
\end{eqnarray}

Similarly to the case of the classical and quantum equations the cancelation of the mixed terms permits the separation of variables,
\begin{equation}
S(\rho,\{\chi_a\})=\omega\rho+\sum_a S(\omega,\chi_a),
\end{equation}
that corresponds to the separation of variables in the quantum equation,
\begin{equation}
\Psi(\rho,\{\chi_a\})=e^{\frac{i}{\hbar}\omega\rho}\prod_a\Psi_a(\chi_a)\sim e^{\frac{i}{\hbar}\omega\rho}\prod_a\psi_a(\chi_a) e^{\frac{i}{\hbar}S_a(\chi_a)}.
\end{equation}

Because $\omega$ is conserved in the classical case these quasiclassical solutions are the ones that correspond to the correct classical limit.

The Hamilton-Jacobi equation on $S_a$,
\begin{equation}
-\frac{\kappa^2}{12}\omega^2(n^{-1}+C_a)-\kappa^2\omega(\partial_aS_a)+\frac{D_a}{2}(\partial_aS_a)^2+V_ae^{\chi_a}=0,
\end{equation}
has exact solution,
\begin{equation}
S_a=\frac{\kappa^2}{D_a}\omega(\chi_a-\chi^{(0)}_a)+\epsilon_a\frac{4P_a}{D_a}\left(\sqrt{1-\frac{V_aD_a}{2P_a^2}e^{\chi_a}}-\operatorname{arctanh}\sqrt{1-\frac{V_aD_a}{2P_a^2}e^{\chi_a}}\right),\label{WKBaction}
\end{equation}
where $\epsilon_a=\pm 1$ and $P_a^2=\frac{\kappa^2\omega^2}{24}(6\kappa^2+D_an^{-1}+C_aD_a)$ and the constant is chosen for solution to vanish at the turning point $e^{\chi^{(0)}_a}=\frac{2P_a^2}{V_aD_a}$.

Treating the action \eqref{WKBaction} as the classical Hamilton-Jacobi action one can obtain the classical trajectories by differentiating it with respect to parameters $\omega$ and $C_a$. Then demanding that,
\begin{equation}
\frac{\partial S}{\partial\omega}=\operatorname{const},\quad \frac{\partial S}{\partial C_a}\Big\vert_{\sum\limits_a C_a=0}=\operatorname{const},
\end{equation}
should yield the equations on the trajectory. Substituting the solutions \eqref{classsol1},\eqref{classsol2} and differentiating with respect to time one obtains,
\begin{eqnarray}
&&-\Big(\frac{P_a}{12}\Big)^2\Big(1-\operatorname{sgn}(P_at-Q_a)\Big)=0,\\
&&-4\epsilon_a\operatorname{sgn}(P_at-Q_a)+4\epsilon_1\operatorname{sgn}(P_1t-Q_1)=0.
\end{eqnarray}
Thus the sign of $\epsilon_a$ is chosen depending on the direction of motion towards the turning point.

The leading order of the prefactor $\psi_a$ happens to be equal,
\begin{equation}
\psi_a=\Big(1-\frac{V_aD_a}{2P_a^2} e^{\chi_a}\Big)^{-1/4}+O(\hbar).
\end{equation}
In the turning point WKB approximation ceases to work however one can match the solutions in the classically forbidden and allowed regions approximating the potential with the linear function $V e^{\chi_a}\sim V_a\chi_a$ and the solutions with Airy function \cite{Bender}. For $\frac{P_a^2}{D_aV_a}>0$ and the solution decreasing at $\chi_a\rightarrow+\infty$ it yields,
\begin{eqnarray}
\Psi_a^{(\chi_a<\chi^{(0)}_a)}\sim \frac{2B}{F^\frac{1}{4}}e^{i\frac{\kappa^2}{\hbar D_a}\omega(\chi_a-\chi^{(0)}_a)}\sin{\frac{4P_a}{\hbar D_a}\left(\sqrt{F}-\operatorname{arctanh}\sqrt{F}+\frac{\pi}{4}\right)},\\
\Psi_a^{(\chi_a>\chi^{(0)}_a)}\sim \frac{B}{(-F)^\frac{1}{4}}e^{i\frac{\kappa^2}{\hbar D_a}\omega(\chi_a-\chi^{(0)}_a)}\exp\left(-\frac{4P_a}{\hbar D_a}\left(\sqrt{-F}-\arctan\sqrt{-F}\right)\right),
\end{eqnarray}
where $B$ is an arbitrary constant and,
\begin{equation}
F=1-\frac{V_aD_a}{2P_a^2}e^{\chi_a}.
\end{equation}
Comparing this result with the asymptotics for Bessel function at large values of order \cite{Bateman} we obtain that it corresponds to the correct asymptotics of $\Psi^{(+)}$ in \eqref{WdWexactsol} if,
\begin{equation}
B=\frac{1}{2}\sqrt{\frac{\hbar\pi D_a}{2P_a}}\exp\Big(i\frac{\kappa^2}{\hbar D_a}\omega\chi^{(0)}_a-\frac{2\pi P_a}{\hbar D_a}\Big)
\end{equation}

In the case of $\frac{P_a^2}{D_aV_a}<0$ we have to choose the solution decreasing at $\chi_a\rightarrow-\infty$,
\begin{eqnarray}
\Psi_a^{(\chi_a>\chi^{(0)}_a)}\sim \frac{2B}{(-F)^\frac{1}{4}}e^{i\frac{\kappa^2}{\hbar D_a}\omega(\chi_a-\chi^{(0)}_a)}\sin{\frac{4\tilde{P}_a}{\hbar D_a}\left(\sqrt{-F}-\arctan\sqrt{-F}+\frac{\pi}{4}\right)},\\
\Psi_a^{(\chi_a<\chi^{(0)}_a)}\sim \frac{B}{F^\frac{1}{4}}e^{i\frac{\kappa^2}{\hbar D_a}\omega(\chi_a-\chi^{(0)}_a)}\exp\left(\frac{4\tilde{P}_a}{\hbar D_a}\left(\sqrt{F}-\operatorname{arctanh}\sqrt{F}\right)\right),
\end{eqnarray}
The following value of $B$ corresponds to the asymptotics of $\Psi^{(+)}$ in \eqref{WdWexactsol2} (using 8.452 in \cite{Gradshtein}),
\begin{equation}
B=\frac{1}{2}\sqrt{\frac{\hbar D_a}{2\pi \tilde{P}_a}}\exp\Big(i\frac{\kappa^2}{\hbar D_a}\omega\chi^{(0)}_a\Big)
\end{equation}

\section{Conclusion: prospects}
We have proposed the model of several scalar fields interacting with gravity that admits exact classical solutions thanks to the separation of variables in the special gauge (for special time variable choice). As well analytical solutions can be found after its quantization in the ADM approach for WdW equation.

The quasiclassical correspondence of the solutions is investigated for the further definition of the physical time. In this paper we considered the special class of scalar field interactions both with positive and negative kinetic terms and potentials but only admitting real classical trajectories. Though all the classical solutions are found the question of stability for partially phantom systems (quintoms \cite{quintom}) and negative potentials unbounded from below is left for further analysis.

In the future we hope to investigate the classical solutions for PT symmetric complex potentials \cite{ptsymmetry} and compare the cosmological trajectories in these potentials with the trajectories in the systems with phantom scalar fields (quintoms \cite{quintom}).

\section{Acknowledgements}The work is done with financial support by Grand RFBR, project 13-02-00127 and by the Saint Petersburg State University grant 11.38.660.2013.

\end{document}